\documentclass[twocolumn,dvipsnames, usenames]{aastex63}
\usepackage{aas_macros}
\usepackage{graphicx}
\usepackage{dcolumn}
\usepackage{bm}
\usepackage{amsmath}
\usepackage{float}
\usepackage{lipsum}
\usepackage{xcolor}
\usepackage{textcomp}
\usepackage{latexsym}
\usepackage{mathtools}
\usepackage{url}
\usepackage{comment}
\usepackage[utf8]{inputenc}
\usepackage[normalem]{ulem}
\usepackage{xspace}
\usepackage{acronym}

\newcommand{\CIERA}{\affiliation{Center for Interdisciplinary Exploration and Research in Astrophysics (CIERA), Northwestern University, 1800 Sherman Ave, Evanston, IL 60201, USA}}
\newcommand{\NUPA}{\affiliation{Department of Physics and Astronomy, Northwestern University, 2145 Sheridan Road, Evanston, IL 60201, USA}}
\newcommand{\Princeton}{\affiliation{Department of Physics, Princeton University, Princeton, NJ 08544, USA}}

\acrodef{GW}{gravitational-wave}
\acrodef{LVK}{LIGO-Virgo-KAGRA}
\acrodef{SNR}{signal-to-noise ratio}
\acrodef{L1}{LIGO-Livingston}
\acrodef{H1}{LIGO-Hanford}
\acrodef{EoS}{Equation of State}
\acrodef{SGWB}{stochastic gravitational-wave background}
\acrodef{TBS}{templated background search}
\acrodef{MC}{Monte Carlo}
\acrodef{BBH}{binary black hole}
\acrodef{BNS}{binary neutron star}
\acrodef{NSBH}{neutron star--black hole}
\acrodef{GMM}{Gaussian mixture model}
\acrodef{BH}{black hole}
\acrodef{NS}{neutron star}
\acrodef{KDE}{kernel density estimate}
\acrodef{O4}{fourth observing run}

\begin{document}
\title{Probing the peak of star formation with the stochastic background of binary black hole mergers}
\correspondingauthor{Sylvia Biscoveanu}
 \author[0009-0000-5584-0320]{Nico Bers}\CIERA\NUPA
\author[0000-0001-7616-7366]{Sylvia Biscoveanu}\email{sbisco@princeton.edu}\thanks{NASA Einstein Fellow}\CIERA\Princeton

\begin{abstract}
    Although the LIGO-Virgo-KAGRA collaboration detects many individually resolvable gravitational-wave events from binary black hole mergers, those that are too weak to be identified individually contribute to a stochastic gravitational-wave background. Unlike the standard cross-correlation search for excess correlated power, a Bayesian search method that models the background as a superposition of an unknown number of mergers enables simultaneous inference of the properties of high-redshift binary black hole populations and accelerated detection of the background. In this work, we apply this templated background search method to one day of simulated data at current LIGO Hanford-Livingston detector network sensitivity to determine whether the weakest mergers contribute information to the detection of the background and to the constraint on the merger redshift distribution at high redshifts. 
    We find that the dominant source of information for the detection of the stochastic background comes from mergers with \aclp{SNR} just below the individual detection threshold. However, we demonstrate that the weakest mergers do contribute to the constraint on the shape of the redshift distribution not only beyond the peak of star formation, but also beyond the redshifts accessible with individually detectable sources.
\end{abstract}

\section{Introduction}
The growing catalog of gravitational-wave events from merging compact-object binaries detected by the \ac{LVK} collaboration includes nearly 300 \ac{BBH} mergers, with over 200 such events observed during the ongoing \ac{O4}~\citep{KAGRA:2021vkt, GraceDB}. These observations have revealed that stellar-mass binary black holes have component masses spanning the range $\sim 2\text{--}100~M_{\odot}$~\citep{Fishbach:2017zga, Talbot:2018cva, KAGRA:2021duu, Tiwari:2020otp, Edelman:2021zkw, Rinaldi:2021bhm}, generally small spins~\citep{Biscoveanu:2020are, Callister:2022qwb, Tong:2022iws, Mould:2022xeu, Galaudage:2021rkt, Roulet:2021hcu}, and a merger rate distribution that is consistent with tracing the cosmic star formation history~\citep{KAGRA:2021duu, Fishbach:2021yvy, Edelman:2022ydv, Payne:2022xan, Ray:2023upk, Callister:2023tgi}. Constraints on the population-level properties of these systems provide clues to their formation and evolution. However, the current \ac{LVK} detectors can only individually resolve \ac{BBH} mergers out to $z\lesssim2$~\citep{Aasi:2013wya}, which limits our ability to probe the properties of black holes at earlier cosmic epochs and to learn about how these properties evolve over time.

In addition to these individually-detectable sources, the \ac{LVK} detectors~\citep{TheLIGOScientific:2014jea, TheVirgo:2014hva, Aso:2013eba, Somiya:2011np, KAGRA:2020tym} are sensitive to a \ac{SGWB} originating from the superposition of unresolvable compact-object mergers. The \ac{SGWB} is typically characterized by the dimensionless \ac{GW} energy density spectrum
\begin{align}
    \Omega_{\mathrm{GW}}(f) \equiv \frac{d \rho_{\mathrm{GW}}}{\rho_{c} d\ln f},
\end{align}
where $\rho_{\mathrm{GW}}$ is the \ac{GW} energy density between $f$ and $f + df$, and $\rho_{c}$ is the critical energy density needed to close the universe. The current best upper limit on $\Omega_{\mathrm{GW}}$ in the band of ground-based \ac{GW} detectors is $\Omega_{\mathrm{GW}}(25~\mathrm{Hz}) \leq 3.4 \times 10^{-9}$ for a spectrum described by $\Omega_{\mathrm{GW}}(f) = \Omega_{\mathrm{ref}}(f/f_{\mathrm{ref}})^{2/3}$~\citep{KAGRA:2021kbb}, where $\alpha=2/3$ is the power-law index expected for a background of unresolved compact-object mergers in the frequency range up to $\mathcal{O}(100)~\mathrm{Hz}$~\citep{Rosado:2011kv, Regimbau:2011rp, Zhu:2011bd, LIGOScientific:2017zlf}. The detection and characterization of this background would allow us to learn about the properties of high-redshift sources inaccessible individually with current detector sensitivities.

Standard searches for the \ac{SGWB} assume the background is Gaussian, stationary, isotropic, and unpolarized, although the latter three assumptions have been relaxed in previous analyses of \ac{LVK} data~\citep[e.g.,][]{Callister:2017ocg, LIGOScientific:2018czr, Mukherjee:2019oma, KAGRA:2021mth, Martinovic:2021hzy}. The optimal search for a signal described by these statistical assumptions involves cross-correlating the data from detector pairs that are sufficiently geographically separated to mitigate common sources of noise (although correlated sources of noise persist at large spatial separations; \citealt{Thrane:2013npa, Thrane:2014yza, Coughlin:2016vor, Meyers:2020qrb, Himemoto:2019iwd, Janssens:2022tdj, Himemoto:2023keu}), such that the signal manifests as excess coherent power~\citep{Christensen:1992wi, Allen:1997ad, Romano:2016dpx, Matas:2020roi}. However, the astrophysical \ac{SGWB} from sub-threshold \ac{BBH} mergers is not expected to satisfy the central limit theorem required for the assumption of Gaussianity, as the duration of these signals in the sensitive band of the \ac{LVK} detectors is $\mathcal{O}(10)~\mathrm{s}$, while the average time between mergers is $\mathcal{O}(100)~\mathrm{s}$, instead producing a ``popcorn-like'' Poisson background~\citep{LIGOScientific:2017zlf}. Thus, the standard cross-correlation search is suboptimal for the \ac{BBH} background, leading to reduced search sensitivity and longer time-to-detection.

Several alternative search methods relaxing the assumption of Gaussianity have been proposed. While the intermittent search method proposed in \citet{Drasco:2002yd, Lawrence:2023buo} models the background as a series of Gaussian ``bursts,'' the Bayesian \ac{TBS} described in \citet{Thrane:2013kb, Smith:2017vfk} models the background as a series of individual compact-object mergers represented by a deterministic signal model that depends on the properties of the merger. By splitting the interferometer strain data into short segments, both searches seek to measure the duty cycle, $\xi$, which represents the fraction of analyzed data segments that contain a merger or the probability that any given segment contains a merger. The intermittent search is found to out-perform the standard cross-correlation search, reducing the time-to-detection for a \ac{BBH} background by a factor of $\sim 50$~\citep{Lawrence:2023buo}. By additionally taking into account the deterministic nature of the signal rather than modeling it as a stochastic process, the \ac{TBS} is found to reduce the time-to-detection by a factor of $\mathcal{O}(1000)$~\citep{Smith:2017vfk}. The \ac{TBS} represents the statistically optimal search for the \ac{SGWB} from \ac{BBH} mergers, at increased computational cost compared to the intermittent search, as full Bayesian inference is performed on each analyzed segment of data to obtain posterior probability distributions on the binary parameters characterizing a potential \ac{BBH} signal in that stretch of data. 

This allows for the simultaneous characterization of the population properties of the \ac{BBH} sources contributing to the background along with the measurement of the duty cycle, another advantage of modeling the \ac{BBH} background deterministically. Previous works have demonstrated that constraints on \ac{BBH} population properties, such as their mass, redshift, and time delay distributions, could be extracted from the point estimate for $\Omega_{\mathrm{GW}}(f)$ obtained using variations of the cross-correlation search~\citep{Zhu:2011bd, Mandic:2012pj, Callister:2016ewt, Safarzadeh:2020qru, Mukherjee:2021ags, Sah:2023bgr, DeLillo:2023srz, Kou:2024gvp}. However, this is necessarily a statistically sub-optimal analysis, as the constraints on these properties come from their imprint on the aggregate $\Omega_{\mathrm{GW}}(f)$ spectrum produced by that \ac{BBH} population~\citep[e.g.,][]{Zhu:2012xw, Renzini:2024pxt} rather than from combining the measurements of the properties of each individual \ac{BBH} event using hierarchical inference~\citep[e.g.,][]{Thrane:2019pe}. A hybrid approach has also been demonstrated that combines the constraints on the \ac{BBH} population properties obtained from resolved signals with those obtained from the (non)detection of the \ac{SGWB} using the cross-correlation search~\citep{Callister:2020arv, Lalleman:2023gul, Turbang:2023tjk,Cousins:2025bas, Ferraiuolo:2025evh}, but this method requires the arbitrary separation of \ac{BBH} mergers into foreground and background.

Instead, the \ac{TBS} allows for simultaneous detection and hierarchical inference of \ac{BBH} population properties without imposing arbitrary thresholds on detection significance of individual events (see also \citealt{Gaebel:2018poe, Farr:2013yna, Messenger:2012jy}). This avoids the issue of accounting for selection effects in hierarchical analyses~\citep{Loredo:2004nn, Mandel:2018mve, 2022hgwa.bookE..45V}, which are often a significant source of systematic uncertainty~\citep[e.g.,][]{Talbot:2023pex, Essick:2023upv}. %
\cite{Smith:2020lkj} demonstrated that informative constraints can be obtained on the \ac{BBH} mass and spin distributions using one week of data, but did not attempt to fit the redshift distribution. The ability to constrain the shape of the merger redshift distribution beyond the detection redshifts of individual events would imply that the most distant mergers contribute information to the detection and characterization of the background. However, the recent result of \citet{Renzini:2024hiu} using a Fisher matrix approach to conclude that the \ac{TBS} obtains 99\% of its information from merger redshifts $z < 1$ casts into doubt the ability of this method to shed light on the properties of high-redshift sources.

In this work, we instead take a Bayesian approach to determine where the \ac{TBS} gets its information, modifying the original likelihood used in \cite{Smith:2017vfk} to incorporate selection effects (above or below a given threshold) to systematically exclude certain sources from the analysis in an unbiased way in Section~\ref{sec:tbs}. We also extend the threshold-less hierarchical inference analysis of \cite{Smith:2020lkj} to the \ac{BBH} merger redshift distribution in Section~\ref{sec:redshift_inference}. While we find that the information driving the posterior on $\xi$ is indeed dominated by the segments with the loudest individual events in qualitative agreement with \citet{Renzini:2024hiu}, we are able to obtain a meaningful constraint on the shape of the \ac{BBH} redshift distribution past the peak of star formation where no individual sources are detected. This implies that while the weakest events do not contribute much information to the detection of the \ac{SGWB} using the \ac{TBS}, they do contribute information to the characterization of the underlying \ac{BBH} population producing the background. Further discussion of the results and caveats of our analysis are presented in Section~\ref{sec:conclusions}. The posterior samples for all analyses presented in this work, including individual-event parameter estimation and hierarchical inference, are available on Zenodo~\citep{bers_2025_17540511}.

\section{The Bayesian Templated Background Search}
\label{sec:tbs}
We use the Bayesian \acf{TBS} to measure the duty cycle for a simulated population of \ac{BBH} mergers. To produce our population, we simulate data from the LIGO Hanford and Livingston detectors with the noise power spectral densities anticipated for their fourth observing run in $4~\mathrm{s}$ segments~\citep{O4_psds}. This duration makes it very unlikely that more than one coalescence will occur in a single segment~\citep{Smith:2017vfk}, given that \ac{BBH} mergers occur every $\sim 200~\mathrm{s}$ in the Universe~\citep{LIGOScientific:2017zlf}. We simulate data in the frequency domain with a frequency range of $[20, 1024]~\mathrm{Hz}$: 20520 Gaussian noise segments and 1080 segments with \ac{BBH} merger signals generated with the \textsc{IMRPhenomPv2} waveform approximant~\citep{Hannam:2013oca} in Gaussian noise, meaning the population has an astrophysical duty cycle $\xi=1080/21600=0.05$. To curtail the large computational cost of performing full Bayesian source characterization on each segment of data discussed later in this section and in Sec.~\ref{sec:conclusions}, we choose a duty cycle of $\xi=0.05$, which is likely significantly larger than the value expected astrophysically. However, we will later demonstrate that our ability to constrain the population properties of high-redshift \ac{BBH} depends only on the number of signal segments and is hence independent of the duty cycle. 

Given that \acp{BBH} come from stellar progenitors, we expect that their merger redshift distribution roughly tracks the cosmic star formation history. The luminosity distances of the signals are thus sampled from the Madau-Dickinson distribution, which is parameterized in redshift as~\citep{Madau:2014bja}
\begin{align}
\psi(z|\gamma, \kappa, z_{p}) &= \frac{(1 + z)^\gamma}{1 + \left(\frac{1 + z}{1 + z_{p}}\right)^\kappa}\\
p(z | \gamma, \kappa, z_{p}) &\propto \frac{1}{(1 + z)} \frac{dV_{c}}{dz} \psi(z | \gamma, \kappa, z_{p})
    \label{eq:madau}
\end{align}
where $dV_{c}/dz$ is the differential comoving volume, and $z_{p}=1.9$, $\kappa=5.6$, $\gamma=2.7$. The distribution is composed of a broken power law that peaks near the redshift of the highest star formation rate, $z_{p}$. The slope of the distribution at low redshifts is parameterized by $\gamma$ and at high redshifts by $\kappa$. We sample from $d_{L} \in [1, 29\times10^3]~\mathrm{Mpc}$, corresponding to a maximum redshift of $z\approx 3.28$. The chirp masses are sampled from a power law distribution, $p(\mathcal{M})\propto\mathcal{M}^{-3.5},\ \mathcal{M} \in [40, 200]M_{\odot}$, where the inclusion of only high masses ensures that a full signal will fit entirely within our segment duration. The spin prior distribution is isotropic and uniform in magnitude, and the mass ratio prior is uniform over $q \in [0.125, 1]$. The shapes of these distributions are chosen to be consistent with current population inference models of \acp{BBH}~\citep{KAGRA:2021duu} and to produce a sufficiently high number of both unresolved and resolved mergers at a manageable computational cost given the size of our simulated dataset, essential for analyzing the information within our population that contributes to the detection of the \ac{SGWB} with the \ac{TBS}.

We perform parameter estimation on all simulated data segments, including both signal and noise segments to obtain evidence estimates and posteriors on all 15 parameters that define each binary, using \textsc{Bilby}, a Python-based Bayesian inference library~\citep{Ashton:2018jfp, Romero-Shaw:2020owr} and the \textsc{Dynesty} sampler~\citep{Speagle:2019dynesty}, analytically marginalizing over the distance and phase parameters~\citep{Thrane:2019pe}. The priors used for parameter estimation are the same as those described above to generate individual \ac{BBH} signals; there is no mismatch between the prior distributions used to simulate the \ac{BBH} population and to infer the parameters of individual events. We use standard priors for the remaining unspecified extrinsic binary parameters~\citep{Romero-Shaw:2020owr}.

Using the signal and noise evidences obtained during sampling for each segment, $\mathcal{Z}_S^i = \mathcal{L}(d_{i} | S)$ and $\mathcal{Z}_N^i = \mathcal{L}(d_{i} | N)$, we calculate the posterior on the duty cycle, $\xi$, the proxy for detecting the \ac{SGWB}. From Bayes' theorem, the posterior is 
\begin{equation}
    p(\xi|\{d\}) = \frac{\mathcal{L}(\{d\}|\xi)\pi(\xi)}{\mathcal{Z}},
\end{equation} where the evidence, $\mathcal{Z}$, is the marginalized likelihood, $\pi(\xi)$ is a uniform prior on the duty cycle for $\xi \in [0,1]$, and the total likelihood of the duty cycle over all $n$ analyzed segments is~\citep{Smith:2017vfk}
\begin{equation}\label{eq:tbs_likelihood}
    \mathcal{L}(\{d\}|\xi)=\prod_i^n(\xi\mathcal{Z}^i_S+(1-\xi)\mathcal{Z}^i_N),
\end{equation}
where the index $i$ indicates the individual data segment.

\begin{figure}
\centering
\includegraphics[width=\columnwidth]{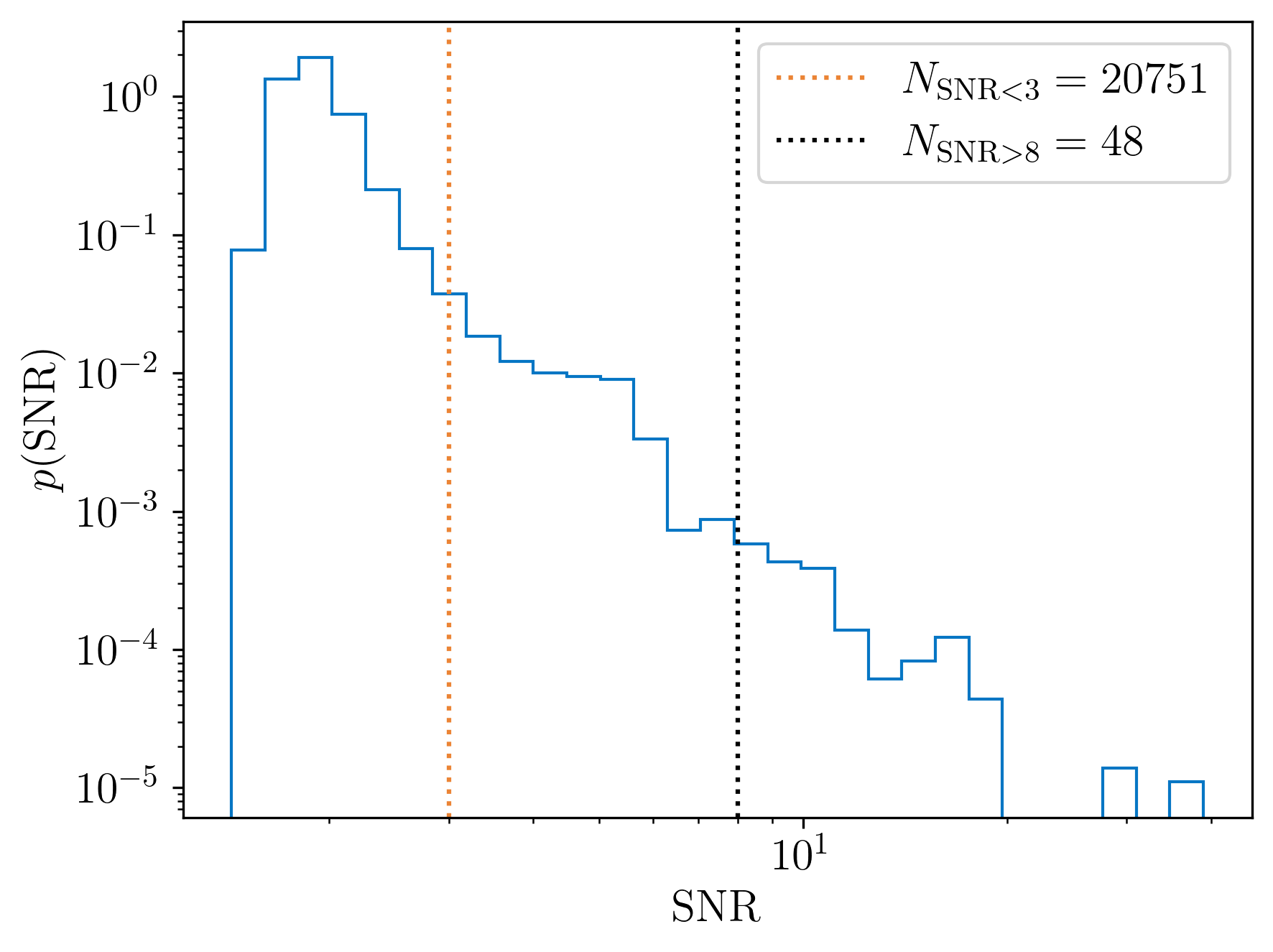}
\caption{Histogram of the median of the posterior on the network matched filter \acp{SNR} for the full dataset with $N=21600$ segments. The dotted vertical orange line at $\mathrm{SNR}=3$ is labeled with the number of segments with \acp{SNR} below this threshold, 20751, which is $\sim96\%$ of the total dataset. The dotted vertical black line at $\mathrm{SNR}=8$ is labeled with the number of segments with \acp{SNR} above this threshold, 48, which is $\sim 0.2\%$ of the total dataset.}
\label{fig:histSNRS}
\includegraphics[width=\columnwidth]{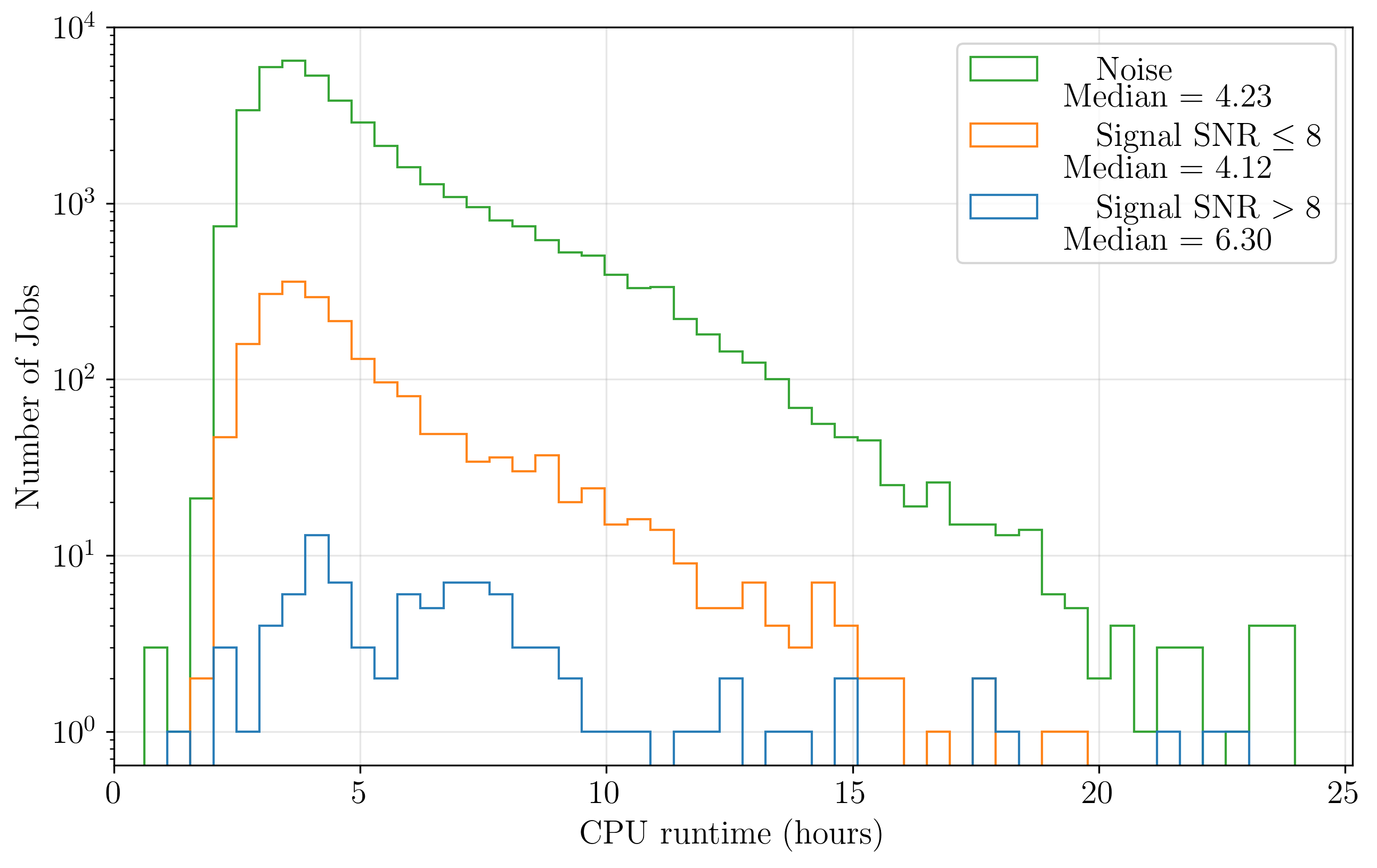}
\caption{Histogram of the CPU runtimes in hours for the per-segment parameter estimation, including two nested sampling runs for each data segment, resulting in $N=43200$ runs. In green is the histogram for noise segments, orange for unresolved signal segments, and blue for resolved signal segments. The total CPU runtime for all the runs is 217034 CPU hours.}
\label{fig:histRuntimes}
\end{figure}

Fig.~\ref{fig:histSNRS} shows the \ac{SNR} distribution for our simulated dataset, including both signal and noise segments. When referring to a \ac{SNR}, unless otherwise specified, we refer to the median of the posterior on the network matched-filter \acp{SNR} obtained during parameter estimation for each segment. The black line in Fig.~\ref{fig:histSNRS} marks what we consider to be ``resolved'' segments with $\mathrm{SNR} > 8$, and thus more than 99.7\% of segments (95.5\% of mergers) are unresolved. The CPU runtime for the individual-segment parameter estimation step is shown in Fig.~\ref{fig:histRuntimes} split by segment category. The median runtimes for unresolved segments are comparable, regardless of whether they contain a signal, while the median runtime for the segments with resolved signals is $\sim50\%$ longer. The total computational cost for two independent nested sampling runs per segment of data (i.e. $N=43200$) is $217034$ CPU hours; we discuss strategies for reducing this runtime in Section~\ref{sec:conclusions}. In the next section, we impose varying cuts based on \ac{SNR} to study how excluding different segments of the population impacts the detection of the \ac{SGWB}, thus varying our definition of a resolved segment. 

\section{Which signal-to-noise ratios contribute information to the TBS?}
\label{sec:information}
\begin{figure*}
\centering
\includegraphics[width=0.45\textwidth]{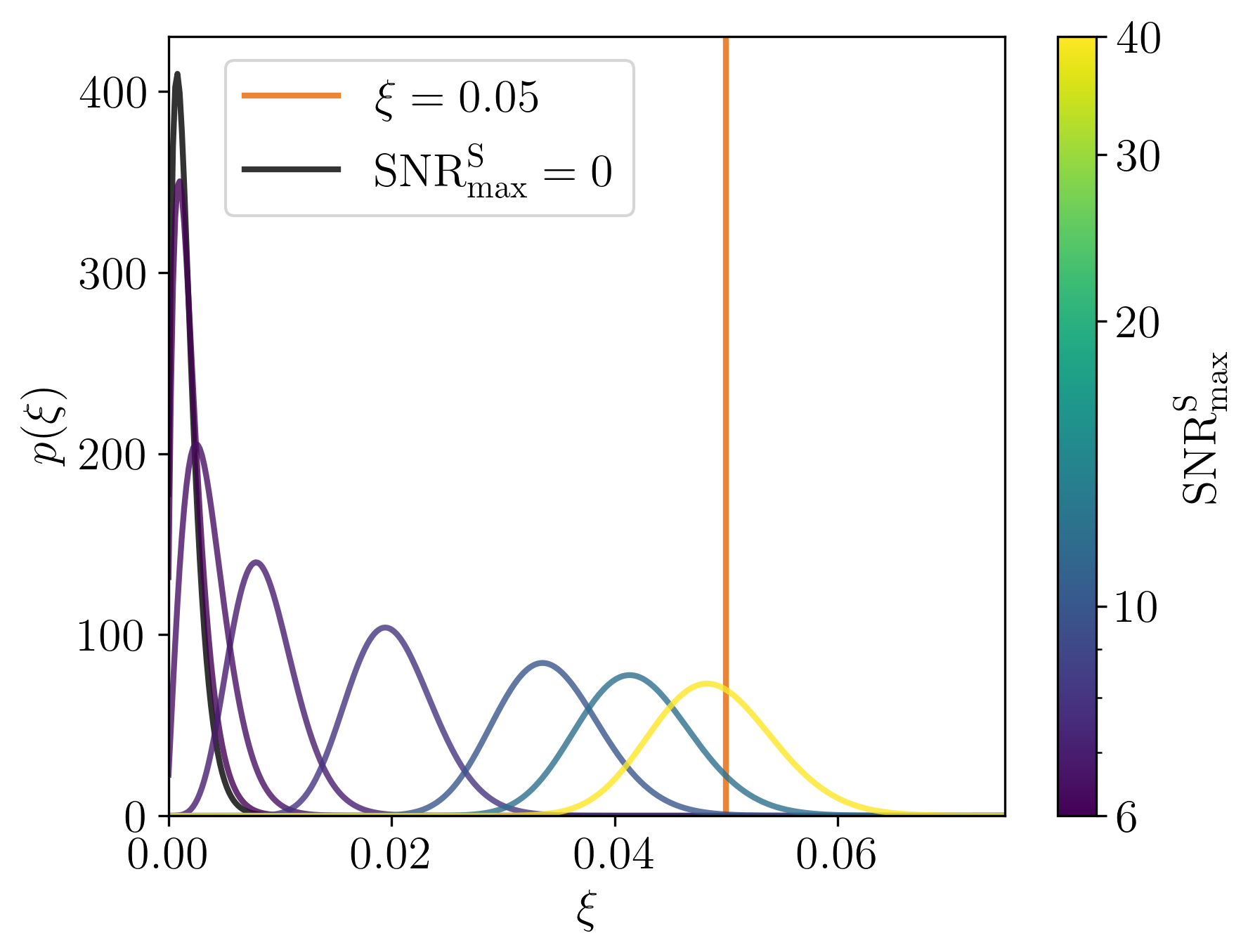}
\includegraphics[width=0.45\textwidth]{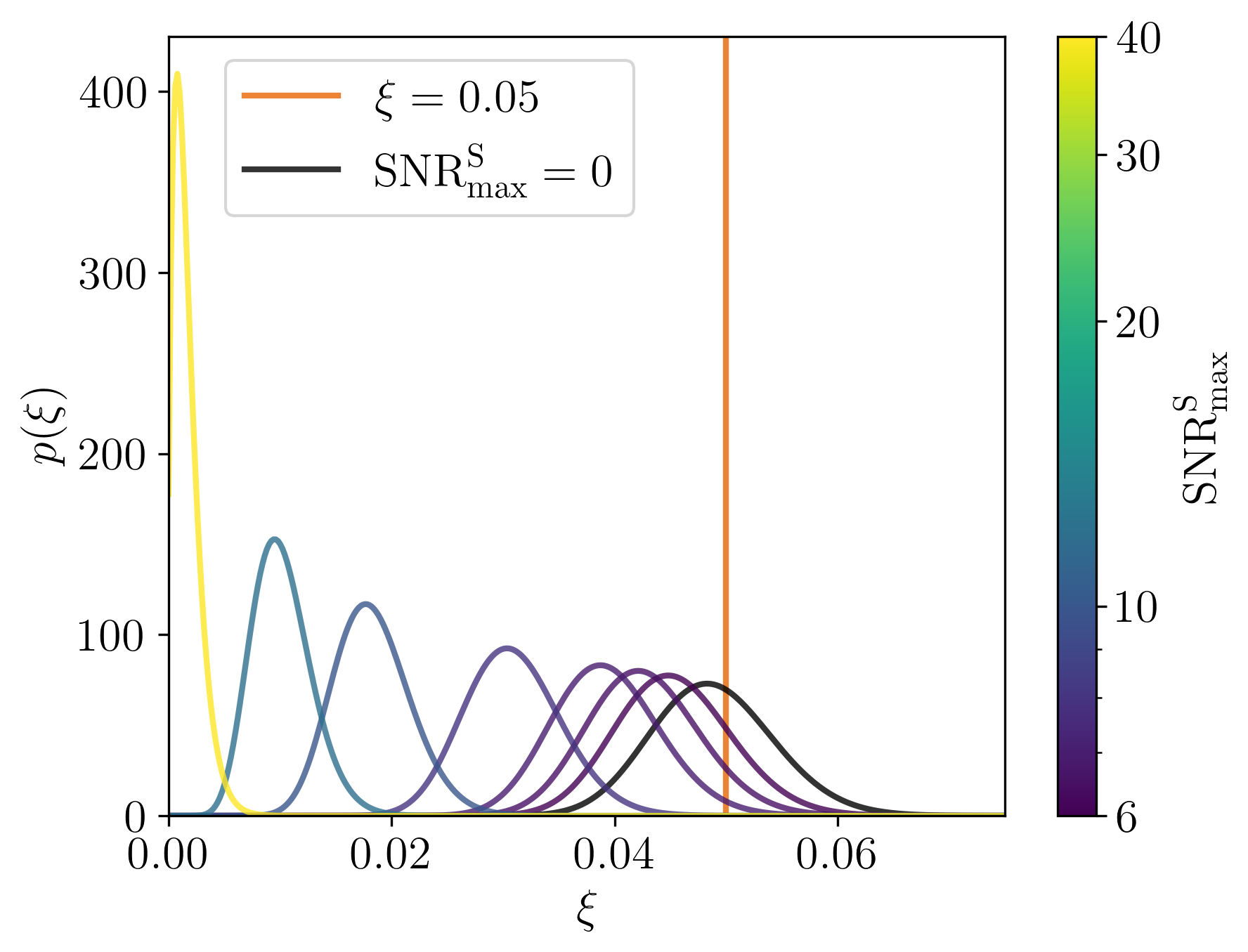}
\caption{\textit{Left:} Posterior on the duty cycle for a population beginning with only noise segments, and replacing them with signal segments with $\mathrm{SNR} < \mathrm{SNR}^\mathrm{S}_{\max}$, for $\mathrm{SNR_{\max}^S} = \{0,~6,~6.5,~7,~8,~10,~12,~40\}$ until $\xi=0.05$ (shown in orange). Note that all signal segments have $\mathrm{SNR} \leq 40$. \textit{Right:} Posterior on the duty cycle beginning with a population with $\xi=0.05$ and replacing signal segments with $\mathrm{SNR} < \mathrm{SNR}^\mathrm{S}_{\max}$ with noise segments until $\xi=0.0$.}
\label{fig:replace}
\end{figure*}

To understand whether the weakest mergers contribute information to the detection of the \ac{SGWB} via the duty cycle constraint obtained through the \ac{TBS}, we calculate the posterior on the duty cycle while replacing noise segments with signal segments and vice versa, in order of ascending \ac{SNR}. By systematically excluding segments of our population based on their \ac{SNR}, we can determine how the exclusion affects the posterior on the duty cycle and thus which merger \acp{SNR} contribute to the information the \ac{TBS} obtains. 
For this part of the study, we down-select 20000 analysis segments from our total population, so that the remaining segments can be used for replacement.
We begin with 20000 noise segments and randomly replace these with signal segments with \acp{SNR} below the variable threshold $\mathrm{SNR_{\max}^S}$. 
The first threshold ($\mathrm{SNR_{\max}^S} = 0$) corresponds to only noise segments included in the population, $\xi=0$, and the final ($\mathrm{SNR_{\max}^S} = 40$) to our true value of $\xi=0.05$, as the highest \ac{SNR} of a segment in the population is $38.93$. At each threshold, we calculate the $\xi$ posterior, as shown in the left panel of Fig.~\ref{fig:replace}. We find that when replacing noise segments with signal segments with $\mathrm{SNR} < 7$, we can measure a non-zero duty cycle with $> 99.99\%$ credibility. Because the posterior first excludes $\xi=0$ with such significance only when segments with \acp{SNR} this high are included in the analysis, we conclude that signals with $\mathrm{SNR} < 7$ do not provide sufficient information to definitively detect the \ac{SGWB} with the \ac{TBS}, at least with this amount of data.

We next perform the same analysis as above, but now beginning with 19000 noise segments and 1000 signal segments for a population with $\xi=0.05$. We replace signal segments with \ac{SNR} below a given threshold with noise segments, until the population contains only noise segments. %
The right panel of Fig.~\ref{fig:replace} shows that when we replace signal segments with $\mathrm{SNR} < 6$ with noise segments, there is still significant support for $\xi=0.05$. Only when we replace signal segments with $\mathrm{SNR} < 7$ out of the population do we measure a duty cycle that excludes $\xi = 0.05$ with $> 97.9\%$ credibility. As we replace signal segments of even higher \acp{SNR} with noise, the posterior peaks closer to $\xi=0$. 

Taken together, these results would suggest that \ac{BBH} mergers with $ \mathrm{SNR} \gtrsim 7$ dominate the information obtained using the \ac{TBS}, and that weaker signals do not contribute any information to the detection of the \ac{SGWB}.
However, the procedure for systematically omitting certain segments from the analysis is not equipped to handle selection thresholds imposed on our population. This means that the posteriors shown in Fig.~\ref{fig:replace} are formally biased. 
We demonstrate how to correct this bias below.

\subsection{Incorporating Selection Effects}
To correct our likelihood calculation for excluding certain events using \ac{SNR} cutoffs, we must take into account selection effects, the probability that a merger with specific binary parameters is detectable.
The likelihood for a single segment given $\xi$ and a detection threshold is given by renormalizing Eq.~\ref{eq:tbs_likelihood}~\citep{Talbot:2021},
\begin{equation}
    \mathcal{L}(d_i|\xi, \mathrm{det}) = \frac{1}{C(\xi)} \left[\xi \mathcal{L}(d_i|S)+(1-\xi) \mathcal{L}(d_i|N)\right].
\end{equation} 
The likelihood must be normalized when integrating over all detectable data realizations,
\begin{align}
    1 &= \int_{d_{\mathrm{det}}}{\mathcal{L}(d_i|\xi, \mathrm{det})}dd, \\
    C(\xi) &= \int_{d_{\mathrm{det}}}\left[\xi \mathcal{L}(d_i|S)+(1-\xi) \mathcal{L}(d_i|N)\right]dd.
\end{align}
We can simplify the right-hand side by defining two new parameters,
\begin{align}
    \alpha = \int_{d_{\mathrm{det}}}\mathcal{L}(d_i|S)dd, \quad
    \beta = \int_{d_{\mathrm{det}}}\mathcal{L}(d_i|N)dd, 
\end{align}
each representing the fraction of astrophysical or noise events that pass a detection threshold~\citep{2022hgwa.bookE..45V}, such that
\begin{align}
    \label{eq:norm_constant}
    \alpha \equiv \frac{N_S^{\uparrow\downarrow}}{N_S},& \quad
    \beta \equiv \frac{N_N^{\uparrow\downarrow}}{N_N},\\
    C(\xi) = \xi\alpha&+(1-\xi)\beta,
\end{align}
where $N_S$ is the number of signal segments, $N_N$ the number of noise segments, and the $\uparrow\downarrow$ superscript denotes whether the cutoff is above or below a \ac{SNR} threshold.
Normalizing Eq.~\ref{eq:tbs_likelihood} by $C(\xi)$,
\begin{equation}
\label{eq:tbs_likelihood_sel_fx}
    \mathcal{L}({d}|\xi, \mathrm{det}) = \prod_i^n\frac1{C(\xi)}[\xi\mathcal{Z}_S^i+(1-\xi)\mathcal{Z}_N^i],
\end{equation}
we obtain the correction to the likelihood accounting for selection bias, allowing us to reanalyze the population with various \ac{SNR} thresholds. 

\begin{figure*}
\centering
    \includegraphics[width=0.7\textwidth]{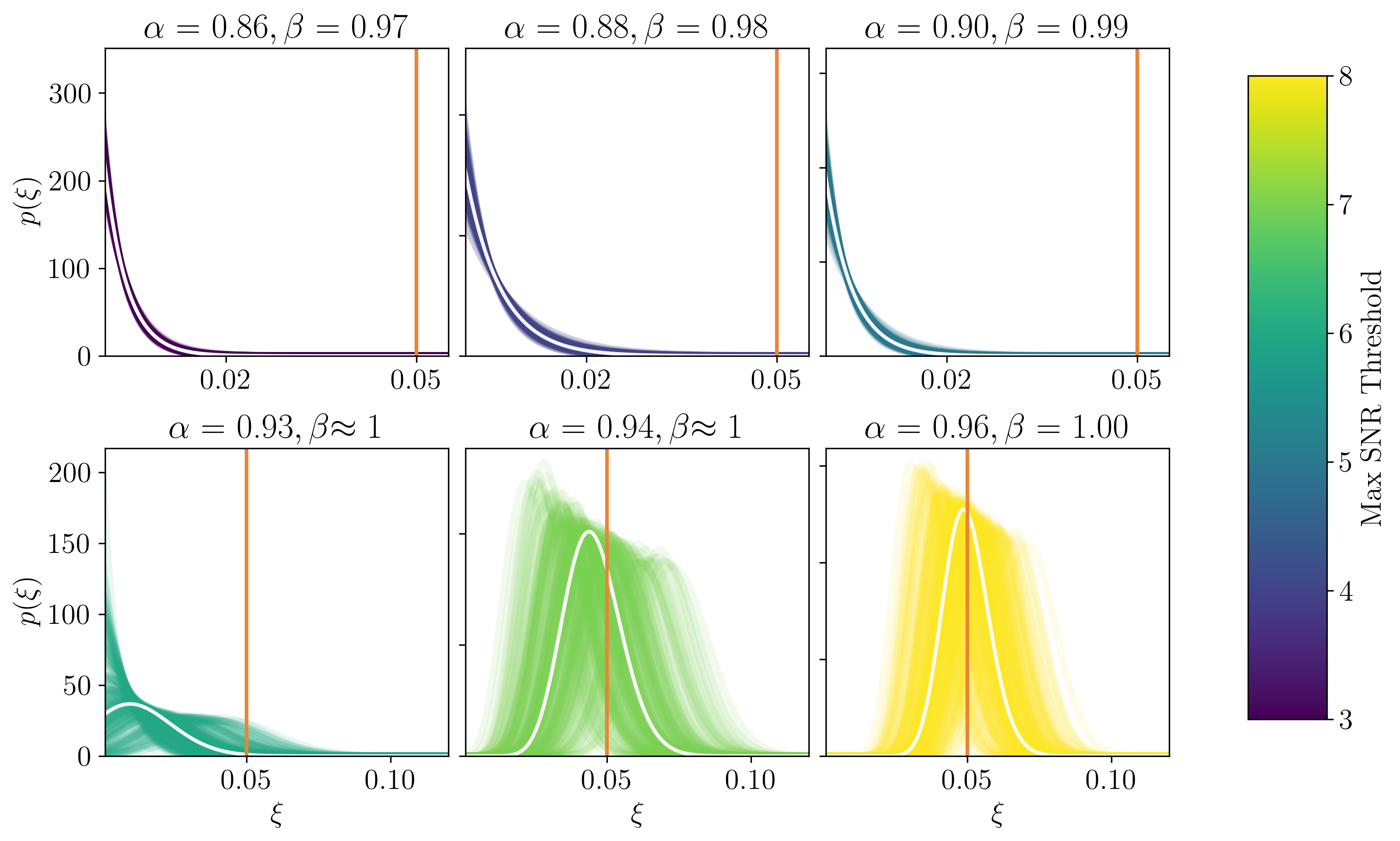}
    \includegraphics[width=0.7\textwidth]{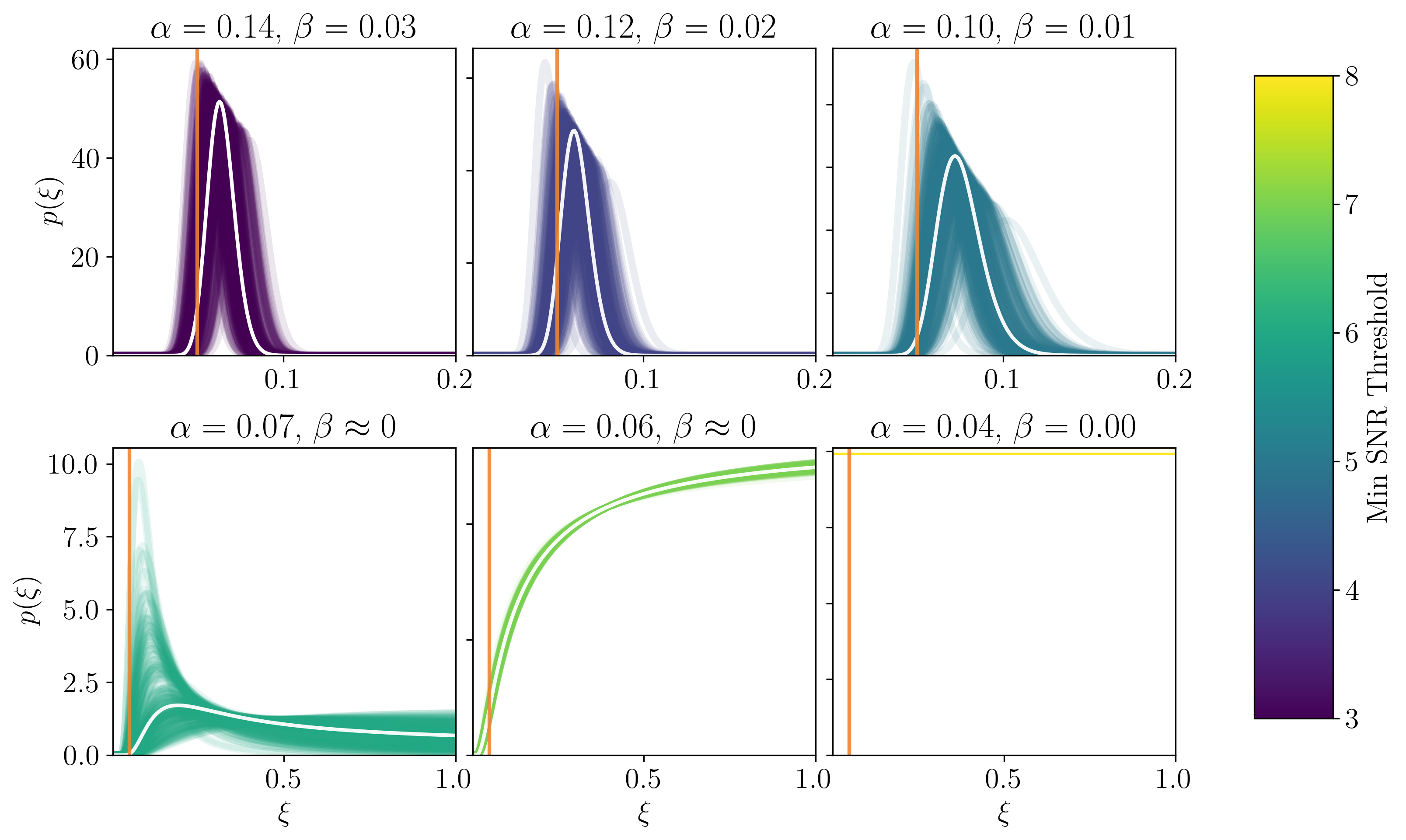}
\caption{Posterior on the duty cycle with selection effects taken into account calculated on the population while only including segments with \acp{SNR} below the maximum \ac{SNR} threshold in the top two rows of the grid, and only including segments with \acp{SNR} above the minimum \ac{SNR} threshold in the bottom two rows of the grid. The colored lines show the posterior on the duty cycle for 200 bootstrapped realization of the dataset while holding the duty cycle of the segments that pass that threshold fixed. The white lines show the posterior calculated with the full population, and the true value of the duty cycle is shown in orange, $\xi = 0.05$. The $\alpha, \beta$ values shown above each plot are the fractions of signal and noise segments, respectively, that pass the selection criteria (see Eq.~\ref{eq:norm_constant}). %
}
\label{fig:selection_effects}
\end{figure*}

We begin by imposing a threshold on the maximum \ac{SNR} of segments included in the analysis, omitting the loudest signals. The top row of Fig.~\ref{fig:selection_effects} shows that when only including segments with $\mathrm{SNR} < 3,~4,~5$, the posterior on the duty cycle peaks at $\xi = 0$, meaning that we do not detect the \ac{SGWB} with only information contained within these sub-threshold populations of \ac{BBH} mergers. However, the $\xi$ posteriors in the second row of Fig.~\ref{fig:selection_effects} with $\mathrm{SNR} < 6,~7,~8$ peak away from $\xi=0$, showing support for the detection of the \ac{SGWB}. 
While the posterior calculated using our full simulated dataset (in white) for $\mathrm{SNR}<6$ peaks away from zero but with only marginal support at $\xi=0.05$, many of the bootstrapped posterior realizations (in color) have significant support at the true value. 

We postulate that this bias in the $\xi$ posterior that remains when excluding segments with $\mathrm{SNR} < 7$ in the analysis even when using the likelihood accounting for selection effects is caused by a subtle breakdown in the statistical assumptions underlying the Bayesian \ac{TBS} method. The uncertainty on the signal and noise evidences obtained by the stochastic sampler likely overwhelms the precision required to distinguish signals from noise at such low \acp{SNR}.
We thus find that there is insufficient information to definitively detect the \ac{SGWB} when only analyzing segments with $\mathrm{SNR}<5$. However, we find growing evidence for a background in a population with segments with only $\mathrm{SNR} < 6$, markedly below current standard detection statistics for individual events corresponding to $\mathrm{SNR}\gtrsim8\text{--}12$. 

Next, we impose a threshold on the minimum \ac{SNR} of segments included in the analysis, omitting the quietest segments. The third row of Fig.~\ref{fig:selection_effects} shows that when we include segments with $\mathrm{SNR} > 3,~4,~5$ in our analysis, we recover the true duty cycle. There is a slight bias towards higher duty cycles in these posteriors, likely because the values of $\alpha$ and $\beta$ in the renormalization for selection effects are calculated from a set of finite \ac{BBH} mergers, meaning there is some inherent uncertainty from the randomness of the merger parameters generated. When we exclude the portion of the population with $\mathrm{SNR} < 6$ (bottom left of Fig.~\ref{fig:selection_effects}), the width of the posterior on $\xi$ increases significantly. The final two posteriors calculated with thresholds of $\mathrm{SNR} > 7,~8$ are extremely broad without definitive peaks because $\beta \sim \mathcal{O}(10^{-4})\approx0$. As discussed above, this is due to the constraints of a finite population size, as given a large enough population, some Gaussian noise would in fact ``pass'' even these high detection thresholds of $\mathrm{SNR} > 7,~8$. These results suggest that the uncertainty in the measurement of the \ac{SGWB} increases as we exclude lower \ac{SNR} events.

\begin{figure}
\centering
\includegraphics[width=\columnwidth]{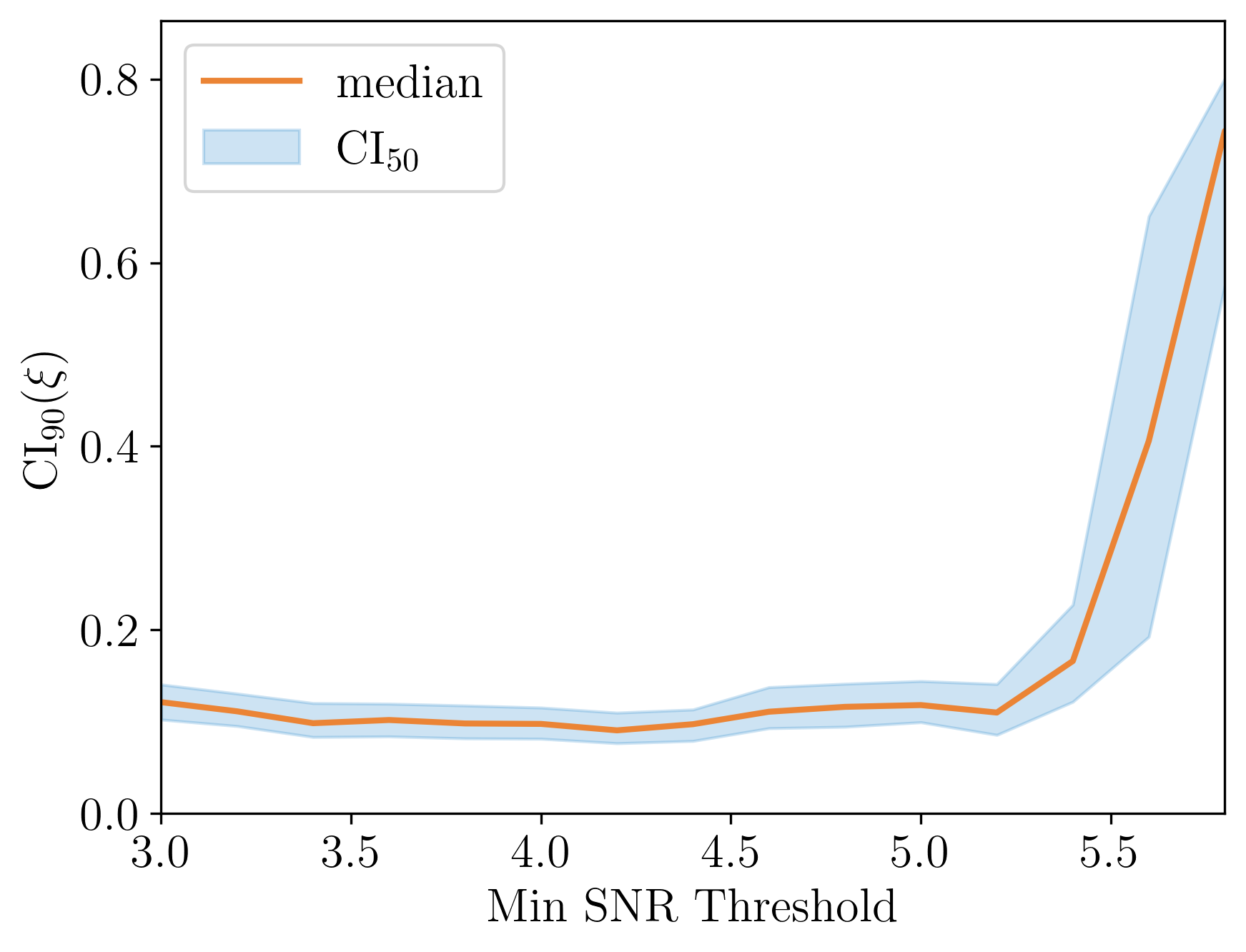}
\caption{Median 90\% credible interval of 200 bootstrapped posterior calculations on $\xi$ in orange, with the shaded blue showing the 50\% credible interval of the median $\mathrm{CI}_{90}$s as a function of the minimum \ac{SNR} threshold imposed. Each bootstrapped posterior is calculated by randomly down-sampling with replacement a set of the segments whose \acp{SNR} are greater than the thresholds such that all posterior calculations include the same number of total segments.}
\label{fig:shadingPlot}
\end{figure}

While we find that the width of the posterior on the duty cycle increases with the minimum \ac{SNR} threshold, this effect could be dominated by the reduced size of the analyzed dataset by imposing increasingly stringent thresholds. To disentangle this effect, we randomly down-sample the full dataset to include the same total number of segments in the analysis, including both signal and noise segments, as the threshold varies. We use $N=54$, corresponding to the number of segments with $\mathrm{SNR} > 7.5$. For each \ac{SNR} threshold, we bootstrap 200 realizations of $N=54$ segments passing the threshold, enforcing that the resulting duty cycle is the same as that obtained by applying that threshold to the full dataset.
We then calculate the $\xi$ posterior and its 90\% credible interval width, $\mathrm{CI}_{90}$, for each realization but note that not all realizations are independent, given the finite size of the dataset.

Fig.~\ref{fig:shadingPlot} shows the $\xi$ posterior $\mathrm{CI}_{90}$ as a function of the minimum \ac{SNR} threshold imposed on the population. The shaded region shows the 50\% credible interval for the calculated $\mathrm{CI}_{90}$ values across the 200 dataset realizations for each threshold. We choose to plot the 50\% credible interval because we only have two independent realizations of $N=54$ segments at the largest \ac{SNR} shown on the x-axis, $\mathrm{SNR}=5.8$, such that
\begin{align}
    \mathrm{CI}_{50} = 1-1/N_{\mathrm{ind}} = 1 - N_{\mathrm{SNR} > 7.5}/N_{\mathrm{SNR} > 5.8}.
\end{align}

While the $\xi$ posteriors remain similarly precise while including segments with $\mathrm{SNR} > 3$ as with $\mathrm{SNR} > 5$, the uncertainty grows by more than a factor of three when we only include segments with $\mathrm{SNR}> 5.8$. Hence, we conclude that the width of the duty cycle posterior increases as segments with $\mathrm{SNR} \approx 5$ are excluded from the analysis, even when keeping the size of the analyzed dataset constant.
This result and results from the selection effects analyses in Fig.~\ref{fig:selection_effects} demonstrate that the dominant information obtained by the \ac{TBS} on the detection of the \ac{SGWB} via the $\xi$ posterior comes from segments with $\mathrm{SNR} \sim 5\text{--} 7$, just below the threshold for individual-event detection, in qualitative agreement with the results of~\cite{Renzini:2024hiu}.

\section{Constraining the \ac{BBH} merger redshift distribution}
\label{sec:redshift_inference}
The results presented in Section~\ref{sec:information} imply that the weakest signals do not contribute significant information to the detection of the \ac{SGWB} with the \ac{TBS},
but we are also interested in determining whether they contribute information to the characterization of the \ac{BBH} population sourcing the background. To this end, we analyze the same population described in Section~\ref{sec:tbs} but now simultaneously infer the duty cycle and the \ac{BBH} population properties, focusing on the redshift distribution. Critically, as of the latest \ac{GW} catalog, GWTC-3~\citep{KAGRA:2021vkt}, no \ac{BBH} mergers have been individually resolved beyond the peak of star formation at $z_{p} \approx  1.9$. Thus, demonstrating a constraint on the merger redshift distribution at larger redshifts would show that we can learn about the properties of \ac{BBH} mergers beyond the peak of star formation without detecting these sources individually at current sensitivities.
\subsection{Hierarchical Bayesian Inference}
We adapt the likelihood in Eq.~\ref{eq:tbs_likelihood} to allow for a flexible population model rather than assuming the fixed model used as the prior for the individual-event parameter estimation~\citep{2020MNRAS.496.3281S}
\begin{equation}
    \mathcal{L}(\{d\}|\Lambda, \xi)=\prod_i^n[\xi\mathcal{L}(d_i|\Lambda, S)+(1-\xi)\mathcal{Z}_{N}^i],
    \label{eq:hyperpe}
\end{equation}
where the likelihood of a single segment of data given the hyper-parameters, $\Lambda$, and the signal hypothesis, $S$, is 
\begin{equation}
    \mathcal{L}(d_i|\Lambda, S)=\int \mathcal{L}(d_i|\theta, S)\pi(\theta|\Lambda)d\theta.
    \label{eq:single_hyper}
\end{equation}

\begin{table}[t]
    \centering
    \begin{tabular}{c c c}
        \hline
        \textbf{Parameter $\Lambda$} & \textbf{True value} & \textbf{Hyper Prior $\pi(\Lambda)$} \\ \hline
        $\xi$ & 0.05 & Uniform(0, 1) \\ 
        $\kappa$ & 5.6 & Uniform(-2, 10) \\ 
        $\gamma$ & 2.7 & Uniform(-2, 10) \\
        $z_{p}$ & 1.9 & Uniform(0.5, 3.28) \\ \hline
    \end{tabular}
    \caption{Hyper-parameter Priors \\ \centering $\xi$ is the astrophysical duty cycle, $\kappa$ describes the slope of the redshift distribution shape at high $z$, $\gamma$ describes the slope of the redshift distribution shape at low $z$, and $z_{p}$ is the peak of redshift distribution.}
    \label{tab:hyper-params}
\end{table}

The likelihood in Eq.~\ref{eq:single_hyper} is typically calculated using \ac{MC} integration to ``recycle'' the binary parameter posteriors obtained under the individual-event parameter estimation prior so that Eq.~\ref{eq:hyperpe} becomes 
\begin{align}
    \mathcal{L}(\{d\}|\Lambda, \xi)=\prod_i^n\left[\frac{\xi}{m}\sum_{j}^{m}\frac{\mathcal{Z}_{S}^i\pi(\theta_{ij}|\Lambda)}{\pi(\theta_{ij} | \Lambda_{0})} 
    +(1-\xi)\mathcal{Z}_{N}^i\right]
    \label{eq:mc_like}
\end{align}
with $\{\kappa, \gamma, z_p\} \in \Lambda$ and $\theta =\ z$ for simultaneous inference of the redshift distribution with the duty cycle. The index $j$ indicates the individual-event posterior sample for $\theta$, and $\Lambda_{0}$ represents the shape of the binary parameter priors used for individual-segment parameter estimation. 
However, we encountered challenges in using the \ac{MC} integration technique due to the size of our population, as the uncertainty in the integral scales with the number of analyzed segments in the population~\citep{Talbot:2023pex}. 

In our population with 21600 total segments, the large uncertainties artificially inflate the likelihood at certain points in parameter space, leading to biased posteriors. We find that $\xi,~\gamma,$ and $z_p$ are biased towards larger values, while $\kappa$ is biased towards smaller values. 
We initially attempted to circumvent this issue by replacing the \ac{MC} integral with a trapezoidal integral, but the various density estimation techniques we used to obtain continuous representations of the individual-event redshift posteriors were insufficiently accurate to obtain unbiased hierarchical inference results for our full population (see Appendix~\ref{appendix:mc_integration} for more details). Instead, we use likelihood reweighting~\citep[e.g.,][]{Payne:2019PhRvD} to map the posterior on $\xi$ given a fiducial population model, $p(\xi | \{d\}, \Lambda_{0})$, to the likelihood given a generic set of hyper-parameters, $\Lambda$, marginalizing over $\xi$:
\begin{align}
    \mathcal{L}(\{d\} | \Lambda) &= \int \mathcal{L}(\{d\} | \Lambda, \xi) \pi(\xi)d\xi, \\
    \mathcal{L}(\{d\} | \Lambda) &= \int \frac{\mathcal{L}(\{d\} | \Lambda, \xi)}{\mathcal{L}(\{d\} | \Lambda_0, \xi)} \mathcal{Z}_0p(\xi | \{d\}, \Lambda_{0})d\xi,
\end{align}
where in the last line we have used Bayes' theorem to replace the product of the likelihood and prior with the product of the posterior and the evidence obtained using the fiducial hyper-parameter values.
Converting the integral into a sum over posterior samples, the total likelihood is
\begin{equation}
    \mathcal{L}(\{d\} | \Lambda)= \frac{\mathcal{Z}_{0}}{\ell}\sum_{k}^{\ell} \frac{\mathcal{L}(\{d\} | \Lambda, \xi_{k})}{\mathcal{L}(\{d\} | \Lambda_0, \xi_{k})}
\end{equation}
where $\mathcal{L}(\{d\} | \Lambda, \xi_{k})$ is given by Eq.~\ref{eq:mc_like}. Once posterior samples are obtained for $\Lambda$, the updated $\xi$ posterior can be reconstructed in post-processing:
\begin{enumerate}
    \item For each posterior sample $\Lambda_{k}$, we calculate $\mathcal{L}_{k}(\{d\} | \xi) = \mathcal{L}(\{d\} | \Lambda_{k}, \xi)$ over a grid of $\xi$ values.
    \item Given the choice of $\xi$ prior assumed during the initial sampling from $p(\xi | \{d\}, \Lambda_{0})$, we calculate the $\xi$ posterior,
    \begin{align}
    p_{k}(\xi | \{d\}) \propto \mathcal{L}_{k}(\{d\} | \xi)\pi(\xi).
    \end{align}
    \item Using inverse transform sampling, we generate a random sample from this posterior associated with that $\Lambda_{k}$ sample.
    \item Repeat this procedure for each $\Lambda_{k}$.
\end{enumerate}

We note that likelihood reweighting is only effective in our use case if the posterior obtained on $\xi$ using the fiducial population model, $p(\xi | \{d\}, \Lambda_0)$, has significant overlapping support with the final posterior obtained under the flexible population model, $p(\xi, \Lambda | \{d\})$. This is easier to guarantee in a simulation, as we chose $\Lambda_0$ to be equal to the true values of $\Lambda$, which is not possible for real data. However, a similar approach could be taken for real data by choosing $\Lambda_0$ to match the maximum likelihood hyper-parameters inferred for the resolved population, which we expect will be similar to the unresolved population, at least at small redshifts. We leave further exploration of the applicability of this likelihood reweighting technique for simultaneous hierarchical inference at scale using real data to future work.

\subsection{Results}
\begin{figure}
\centering
\includegraphics[width=\columnwidth]{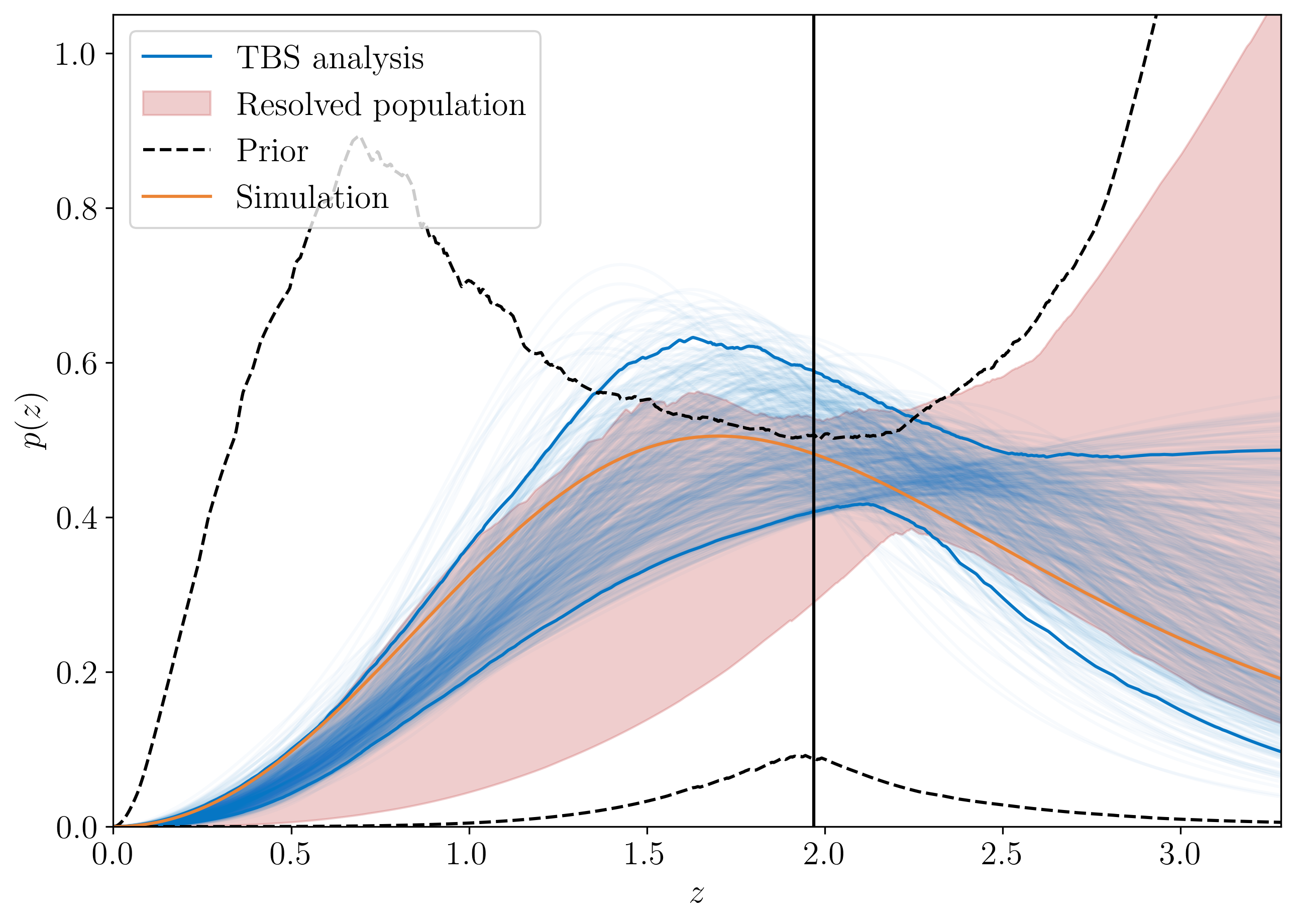}
\caption{Inferred redshift distribution obtained using hierarchical inference. 
In blue thin lines are posteriors on the redshift distribution calculated using the Madau-Dickinson parameterization for individual hyper-parameter posterior samples. The solid blue lines bound the $\mathrm{CI}_{90}$ for the full population analysis, the orange solid line is the true population prior Madau-Dickinson distribution, and the dashed black lines are the $\mathrm{CI}_{90}$ prior bound. The shaded red region is the $\mathrm{CI}_{90}$ for the analysis including only resolved segments and taking into account selection effects, and the solid  black vertical line at $z=1.97$ marks the redshift of the farthest signal in the population with $\mathrm{SNR}> 8$.}
\label{fig:ppdAll}
\end{figure}
Fig.~\ref{fig:ppdAll} shows the posterior on the redshift distribution obtained in the analysis of the entire population compared to the simulated Madau-Dickinson redshift distribution from which the population was drawn. We are able to recover the shape of the simulated redshift distribution of \ac{BBH} mergers out to redshifts beyond those where resolved events are individually detected. Crucially, the Bayesian \ac{TBS} enables us to constrain the posterior on the redshift hyper-parameters significantly not only relative to the prior but also relative to a standard hierarchical analysis limited to the resolved population with $\mathrm{SNR} > 8$ accounting for selection effects with fixed $\xi=1$ (red shading). By not distinguishing between a foreground and background, we find that the unresolved population must be contributing to the constraint on the redshift distribution at high redshifts. The narrowing of the constraint on the shape of the redshift distribution seen at $z\approx2$ is due to the parameterization used, as the prior is also narrower. The parameterization also causes the lower bound of the \ac{TBS} posterior on the redshift distribution to fall below the lower bound of the resolved population posterior at high redshifts ($z\gtrsim2$). The merger rate at large redshifts inferred with the resolved analysis falls as quickly as the parameterized form allows given the increased support for lower merger rates at small redshifts relative to the analysis including unresolved mergers. 

\begin{figure}
\centering
\includegraphics[width=\columnwidth]{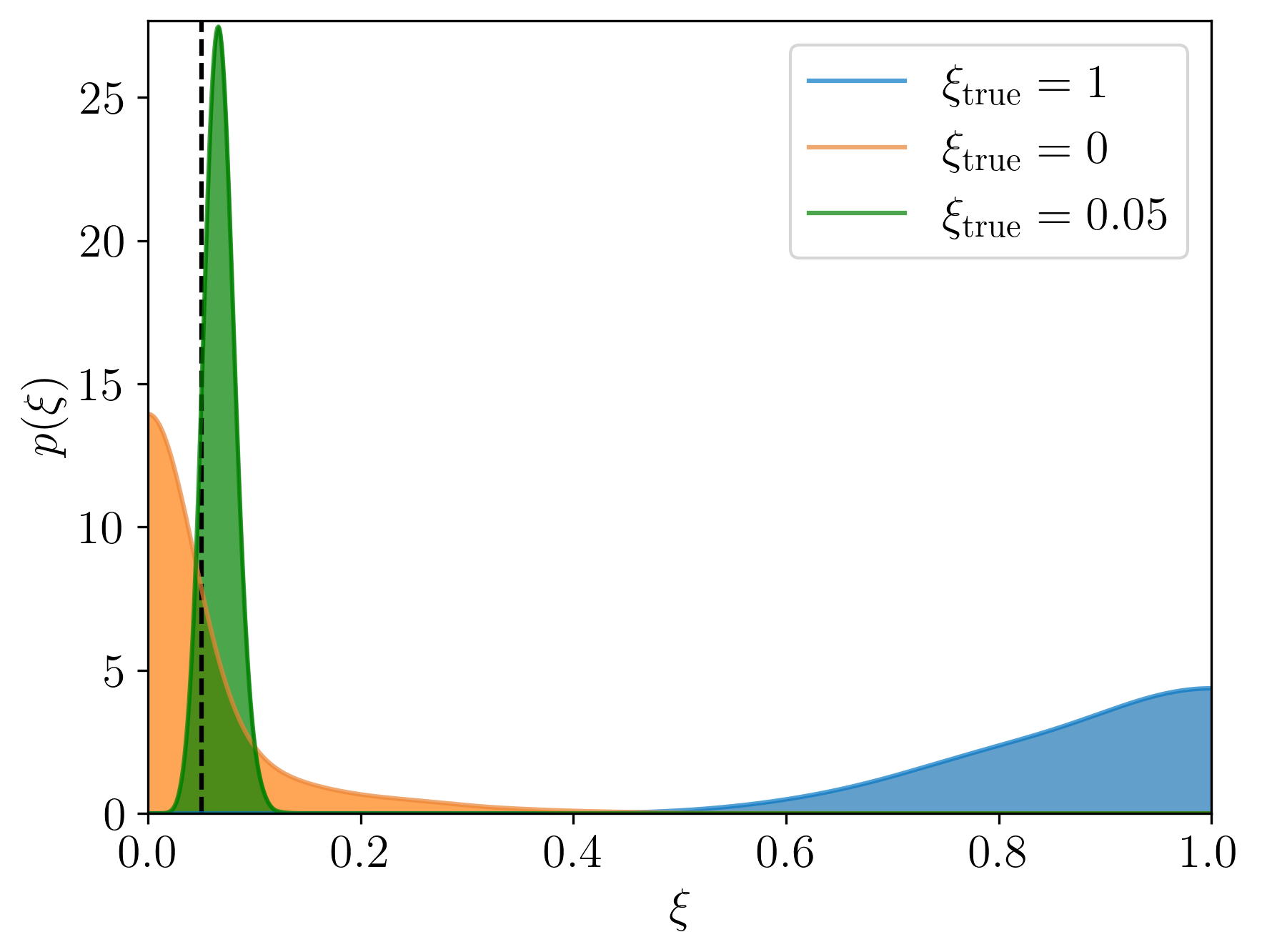}
\caption{Posteriors on the duty cycle from hierarchical inference with all segments in green demonstrating that the \ac{TBS} is unbiased, only noise segments in orange demonstrating that the \ac{TBS} is safe, and only signal segments in blue demonstrating that the \ac{TBS} is effective.}
\label{fig:xiPostCompare}
\end{figure}

When analyzing the simulated data with the above-described likelihood reweighting method for hierarchical inference on the redshift distribution and the duty cycle, we test to ensure that the \ac{TBS} is safe, unbiased, and effective. As shown in the $\xi$ posterior comparison plot in Fig.~\ref{fig:xiPostCompare}, we find that all three are true: When only Gaussian noise segments are analyzed (orange), we find that the posterior on $\xi$ peaks at 0, meaning the method is safe. When the entire population is analyzed (green), we find that the posterior on $\xi$ has significant support at the true value of $\xi=0.05$, meaning the method is unbiased. Finally, when analyzing only the signal subset of the population (blue), we find the posterior peaks at 1.0 and rules out $\xi=0$, meaning the method is effective at detecting the \ac{SGWB}. %

\begin{figure}
\centering
\includegraphics[width=\columnwidth]{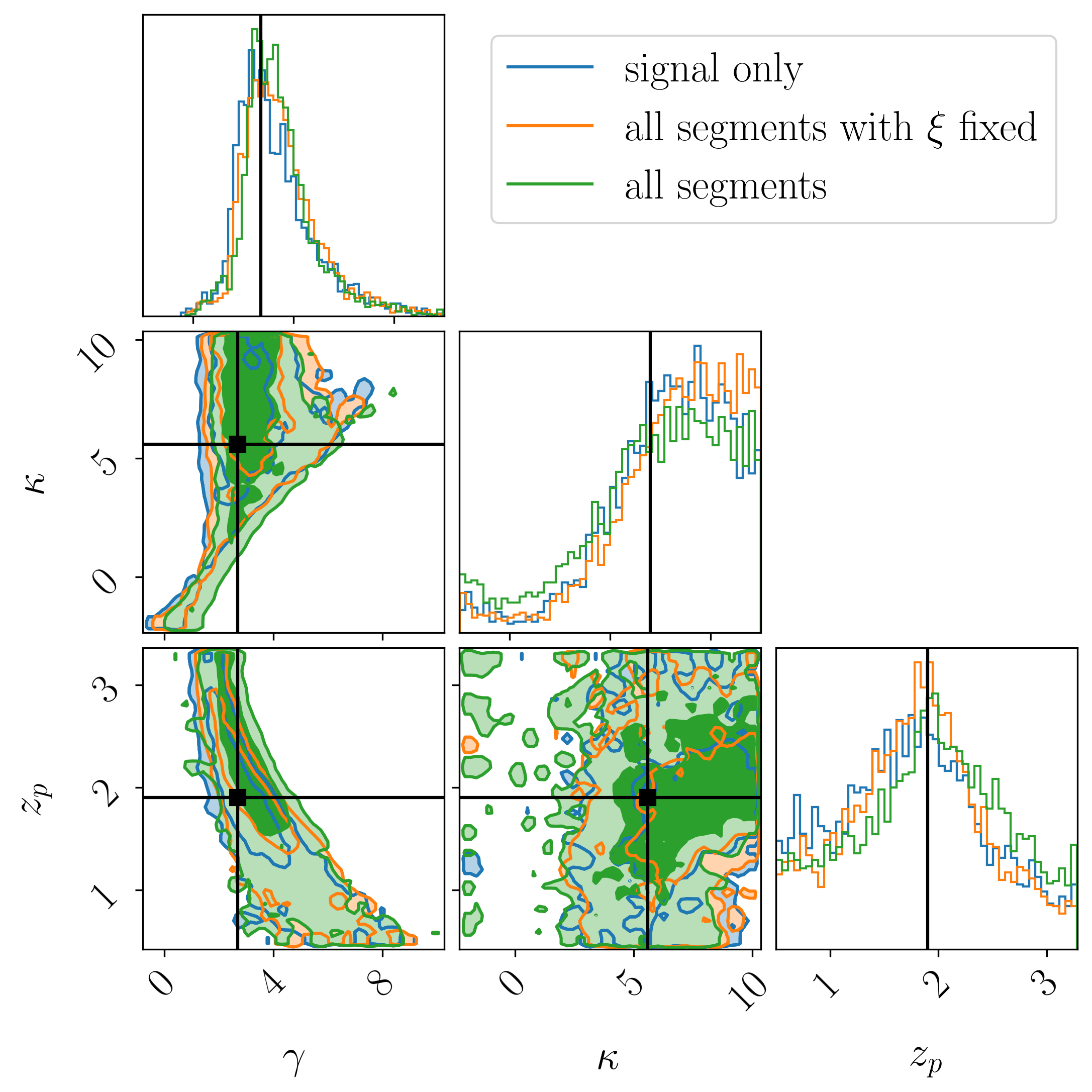}
\caption{Comparison corner plot of the posteriors on the redshift distribution hyper-parameters from hierarchical inference using all segments in green, only signal segments in blue, and all segments with the duty cycle fixed to the true value in orange. The darker region is the 90\% credible region, and the lighter is the 50\% credible region. The black lines indicate the true values of the hyper-parameters.}
\label{fig:cornerCompThree}
\end{figure}

While we have chosen an artificially inflated value of the duty cycle to curtail the computational cost of our simulation, we want to demonstrate the robustness of our constraints on the redshift distribution for a more realistic choice of $\xi$. Fig.~\ref{fig:cornerCompThree} shows that the posteriors for the individual hyper-parameters that describe the redshift distribution are consistent when analyzing just signal segments, i.e., $\xi_{\mathrm{true}}=1$, all segments but fixing $\xi_{\mathrm{true}}=0.05$, and when analyzing the whole population with the method described above. This means that varying the true value of $\xi$ does not affect our ability to constrain the redshift distribution. Imperfect knowledge of $\xi$ (simultaneous inference) similarly does not affect the constraints obtained on the population distribution. As such, the constraints on the redshift distribution that we demonstrate for our simulated population can be obtained with any amount of data required for the detection of 48 resolved, high-mass \acp{BBH} at \ac{O4} sensitivity, whether that corresponds to one day as in our simulation, or more realistically $\mathcal{O}(\mathrm{months})$.

\section{Conclusions}
\label{sec:conclusions}
In this work, we seek to determine where the information driving the detection of the \ac{SGWB} using the Bayesian \ac{TBS} comes from and whether the farthest, unresolved \ac{BBH} mergers contribute to the constraint on the merger redshift distribution at large redshifts beyond the peak of star formation.
We begin by imposing \ac{SNR} thresholds to determine which segments contribute to the constraint on the duty cycle, both ignoring and accounting for selection effects. %
Our results demonstrate that signal segments with \acp{SNR} between $\sim 5\text{--}7$ dominate the information on the detection of the \ac{SGWB} that is accessible with the \ac{TBS}, in qualitative agreement with~\cite{Renzini:2024hiu}, which instead uses a Fisher information matrix to directly calculate contributions to the detection of the \ac{SGWB}. 

We then expand upon the work of~\cite{Smith:2020lkj} to use hierarchical Bayesian inference to constrain the merger redshift distribution simultaneously with the duty cycle. %
With one day of simulated \ac{GW} data at \ac{O4} sensitivity, we show that while weak segments do not contribute significantly to the measurement of the duty cycle obtained with the \ac{TBS}, the individually unresolved segments \textit{do} constrain the posterior on the redshift distribution at redshifts beyond the peak of star formation and beyond where any signals are resolved with current detector sensitivities. The constraints on the redshift distribution are not only informative relative to the prior but also relative to an analogous hierarchical analysis on resolved events only.

In this work, we have focused on high-mass \acp{BBH} to simplify the analysis by ensuring that all signals fit the same segment duration and that our simulated population includes a minimum number of detectable signals without requiring a prohibitively large number of total signal segments to be simulated. For the analysis of a real population including lower-mass signals, a similar approach could be employed---targeting signals in specific mass bins independently---as preliminary studies suggest minimal contamination of the duty cycle measurement from mergers with masses outside the allowed prior range. In this case, the measurement of the duty cycle would provide a proxy for the \ac{SGWB} from \ac{BBH} mergers in each mass range. Our analysis should be interpreted as an upper limit on the constraining power that could be achieved when the results from different mass ranges are combined. Each would provide information about different redshift ranges, as the \ac{SNR} distribution is a strong function of both mass and redshift. We leave the development of a method to include \ac{BBH} mergers with different mass ranges in the \ac{TBS} framework to future work.

While our results also show that the value of the duty cycle does not impact the ability to infer the redshift distribution hyper-parameters, it is important to recognize that our analysis is using a significantly inflated value of $\xi$ compared to the value expected astrophysically given our population~\citep{Smith:2017vfk}. We leave an analysis that uses a more realistic value of $\xi \sim \mathcal{O}(10^{-4})$ to future work in which several changes could be made to the analysis settings to decrease computational costs shown in Fig.~\ref{fig:histRuntimes}. For example, various forms of likelihood acceleration, including multi-banding and relative binning, could be used to lower the compute time~\cite[e.g.,][]{vinciguerra:2017acc,Krishna:2023bug,Zackay:2018qdy,Morisaki:2021ngj}, and sampler settings like the number of nested sampling live points could be tuned to optimize the trade-off between runtime, evidence estimation error, and number of final posterior samples. While the analysis of more data including more signal segments would improve the constraints on the duty cycle and redshift distribution hyper-parameters, informative constraints can be obtained with a subset of the data from a full observing run to curtail the computational cost. The compute time for the hierarchical Bayesian inference step in Sec.~\ref{sec:redshift_inference} is negligible.

We also note that our analysis is performed on simulated frequency-domain data, and there are non-trivial modifications that would be needed to use the \ac{TBS} with real time-domain data. One such factor is the uncertainty in the estimated power spectral density of real data; another is the off-diagonal terms that appear in the likelihood noise covariance matrix that are ignored in frequency-domain simulations due to assumed periodic boundary conditions for the noise that lead to a diagonal matrix~\citep{Talbot:2020auc, Talbot:2021igi, Kou:2025bhk}. 

While we only infer the \ac{BBH} merger redshift distribution, an extension of this work should simultaneously infer the mass and spin distributions alongside the redshift, as their posteriors may be correlated, leading to an increase in the uncertainty in the redshift distribution inference. Additionally, population-level correlations are expected between the \ac{BBH} mass and redshift distributions but have not been found with resolved sources~\citep{LIGOScientific:2025pvj, Lalleman:2025xcs}. \ac{TBS} could expand the redshift horizon, paving the way for future investigations with higher redshift sources, which would require the inclusion of lower mass mergers in the analysis as discussed above.

In addition to fitting both the mass and spin distributions, in future work, we will explore using a non-parametric redshift distribution model such that the priors and assumed parameterization have a smaller impact on the shape of the posterior. As shown in Fig.~\ref{fig:ppdAll}, the prior causes the posteriors on the redshift to narrow right after $z\approx2$, whereas this would not necessarily be the case given a non-parametric redshift distribution model. Finally, an important future extension of our work is a direct comparison of the redshift distribution constraint obtained with the \ac{TBS} to the method described in \cite{Callister:2020arv} that analyzes the foreground and background separately, as opposed to our threshold-less analysis.

\vspace{7pt}
The authors thank Colm Talbot and Vicky Kalogera for insightful suggestions and discussions and Saleem Muhammed for providing helpful comments during internal LVK document review. 
This material is based upon work supported by NSF's LIGO Laboratory which is a major facility fully funded by the National Science Foundation.
LIGO was constructed by the California Institute of Technology and Massachusetts Institute of Technology with funding from the National Science Foundation and operates under cooperative agreement PHY-0757058. 
S.B. is supported by NSF PHY-2513246 and by NASA through the NASA Hubble Fellowship grant HST-HF2-51524.001-A awarded by the Space Telescope Science Institute, which is operated by the Association of Universities for Research in Astronomy, Inc., for NASA, under contract NAS5-26555.
N.B. is supported by NSF grant PHY-2207945 and through a NASA grant awarded to the Illinois/NASA Space Grant Consortium. 
Any opinions, findings, conclusions, or recommendations expressed in this material are those of the authors and do not necessarily reflect the views of NASA. %
The authors are grateful for computational resources provided by the Caltech LIGO Laboratory supported by NSF PHY-0757058 and PHY-0823459. This paper carries LIGO document number LIGO-P2500381.

\bibliography{references}

\appendix
\section{Dealing with Monte Carlo integral uncertainty}
\label{appendix:mc_integration}
With 21600 individual-segment redshift posteriors included in our hierarchical analysis, the uncertainty associated with the \ac{MC} integral in Eq.~\ref{eq:mc_like} becomes untractably large, resulting in spuriously large likelihood values at certain points in the parameter space. In particular, we find that the variance in the \ac{MC} integral is strongly correlated with the inference of the duty cycle and $\gamma$ and anticorrelated with $\kappa$. This leads to a bias towards larger values of $\xi, \gamma$ and small values of $\kappa$, corresponding to redshift distributions that peak strongly at the edge of the allowed redshift range.

We explore several methods to mitigate this bias. The variance in the \ac{MC} integral can be reduced by increasing the number of samples over which the sum is performed. To generate more samples without repeating the computationally-expensive individual-event parameter estimation step, we can use density estimation to obtain a continuous representation for $p(z | d_{i})$ for each segment from which we can draw samples. We first attempt to use a \ac{KDE} as implemented in \textsc{PESummary}~\citep{Hoy:2020vys}, which is well-suited to density estimation problems with low dimensionality. We also attempt to fit $p(z | d_{i})$ using a three-component \ac{GMM} as implemented in \textsc{scikit-learn}, using the CDF mapping proposed in \cite{Talbot:2020oeu} to avoid edge effects due to sharp cutoffs in the posteriors at the upper edge. The number of components is determined empirically by comparing the performance of the \ac{GMM} using a variable number of components on a reserved testing subset of the original posterior samples for a few individual events. The performance of the \ac{GMM} is quantified using the score, or average log-density evaluated over the testing samples.

\begin{figure}
\centering
\includegraphics[width=0.55\columnwidth]{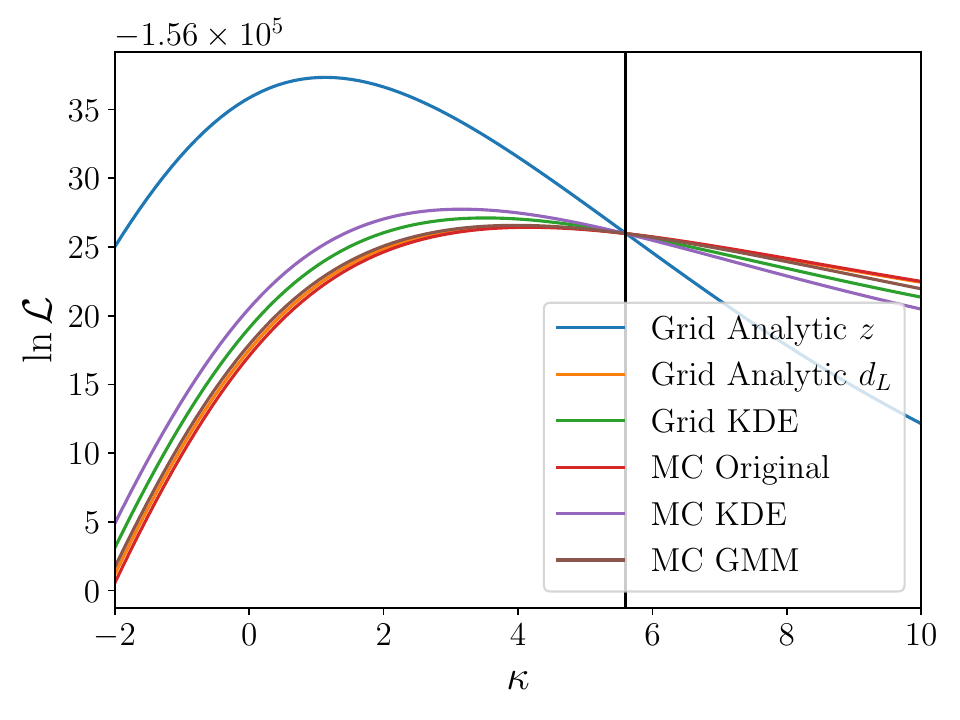}
\caption{Log Likelihood as a function of $\kappa$ for a population consisting of 500 signal segments with $\xi, \gamma, z_{p}$ fixed to their true values for each of the six methods of evaluating Eq.~\ref{eq:hyperpe} that we explored.}
\label{fig:likelihood_comp}
\end{figure}

We then use inverse transform sampling from the posterior density approximated using both the \ac{KDE} and \ac{GMM} methods to generate 20000 posterior samples for each event, roughly an order of magnitude more than were produced by the original nested sampling analysis, feeding these new samples to the \ac{MC} integral in Eq.~\ref{eq:mc_like}. While the \ac{GMM} provides an improved fit compared to the \ac{KDE}, we find that neither method provides a sufficiently accurate representation of the posterior to ensure an unbiased hierarchical inference likelihood. This is demonstrated in Fig.~\ref{fig:likelihood_comp}, which shows the log likelihood as a function of $\kappa$ with the other hyper-parameters fixed to their true values for the various likelihood evaluation methods we explore, applied to a population of 500 signal segments. Compared to the likelihood evaluated using a \ac{MC} integral over the original samples generated by nested sampling shown in red, both the \ac{KDE} and \ac{GMM} \ac{MC} integral likelihood evaluation methods return larger likelihood values at small $\kappa$ values and smaller likelihood values at large $\kappa$, exacerbating the original bias we found using \ac{MC} integration over the original posterior samples for the full population of 21600 segments.

We also attempt to eschew \ac{MC} integration altogether, instead using a trapezoidal integral over a grid of redshifts for Eq.~\ref{eq:single_hyper},
\begin{align}
  \mathcal{L}(d_i|\Lambda)\approx\sum_{j=1}^{M} \frac{\mathcal{L}(d_i|z_{j-1})\pi(z_{j-1}|\Lambda) + \mathcal{L}(d_i|z_j)\pi(z_{j}|\Lambda)}{2} \Delta z_{j}. 
  \label{eq:trapz}
\end{align}
where we use Bayes' theorem to replace the single-event likelihood with the ratio of the posterior and the original parameter estimation prior. While performing this gridded integral over the \ac{KDE} approximation for the individual-event posteriors leads to less hierarchical likelihood bias than the \ac{MC} integral over samples drawn from the \ac{KDE} posterior, the bias is not fully ameliorated. 

We are only interested in the individual-event likelihoods marginalized over all parameters except for redshift, which are directly derived from the luminosity distance likelihoods given a cosmology. Since we analytically marginalize over luminosity distance during sampling, we can directly reconstruct this likelihood without relying on any density estimation technique and pass this gridded likelihood directly into the trapezoidal integral in Eq.~\ref{eq:trapz}. This approach, shown in orange in Fig.~\ref{fig:likelihood_comp}, yields the least bias in the hierarchical likelihood compared to the \ac{MC} integral over the original samples. While we find unbiased hierarchical inference results using this method applied to only the signal segments in our population, the analytic likelihood reconstruction is not sufficiently accurate when we begin incorporating noise segments into our analysis, leading to the same kind of bias we found with the \ac{MC} integral over the original posterior samples applied to our full population.

The most severe bias among the likelihoods shown in Fig.~\ref{fig:likelihood_comp} is introduced by the numerical conversion between the distance and redshift priors performed by \textsc{Bilby}. While we are interested in constraining the population-level redshift distribution, we draw injections and perform individual-event parameter estimation sampling from the corresponding luminosity distance distribution. To generate the equivalent luminosity distance prior from the Madau-Dickinson redshift distribution in Eq.~\ref{eq:madau}, \textsc{Bilby} calculates the Jacobian $\frac{dd_{L}}{dz}$ numerically by taking the gradient between the luminosity distance and redshift arrays calculated using the \textsc{astropy} cosmology module~\citep{Astropy:2022ucr}. We find that this conversion introduces sufficient error so that analytically reconstructing the redshift posterior from the distance likelihood as 
\begin{align}
p(z | d) &\propto \mathcal{L}(d | d_{L}) \frac{dd_{L}}{dz} \pi(z)
\end{align}
leads to significant biases in the hierarchical inference likelihood (blue line in Fig.~\ref{fig:likelihood_comp}), while converting directly from the distance posterior to the redshift posterior (orange line in Fig.~\ref{fig:likelihood_comp}),
\begin{align}
p(z | d) &= p(d_{L} | d)\frac{dd_{L}}{dz},
\end{align}
provides the closest match to the hierarchical likelihood calculated using \ac{MC} integration over the original posterior samples.

\end{document}